\documentclass[conference]{IEEEtran}
\IEEEoverridecommandlockouts

\usepackage{booktabs}
\usepackage{multirow}
\usepackage{array}
\usepackage{amsmath,amssymb}
\usepackage{listings}
\usepackage{xcolor}
\usepackage{caption}
\usepackage{graphicx}
\usepackage{tikz}
\usetikzlibrary{positioning,arrows.meta}
\usepackage{hyperref}
\definecolor{demonlink}{rgb}{0.10,0.20,0.55}
\hypersetup{colorlinks=true,linkcolor=demonlink,citecolor=demonlink,urlcolor=demonlink}

\graphicspath{{./}}

\definecolor{codecomment}{rgb}{0.40,0.45,0.50}
\definecolor{codekeyword}{rgb}{0.10,0.20,0.55}
\definecolor{codestring}{rgb}{0.20,0.45,0.20}

\lstdefinestyle{pseudo}{
  basicstyle=\footnotesize\ttfamily,
  keywordstyle=\color{codekeyword}\bfseries,
  commentstyle=\color{codecomment}\itshape,
  stringstyle=\color{codestring},
  showstringspaces=false,
  keepspaces=true,
  columns=fullflexible,
  breaklines=true,
  language=Python,
}

\lstdefinestyle{ascii}{
  basicstyle=\scriptsize\ttfamily,
  keepspaces=true,
  columns=fullflexible,
  breaklines=false,
}

\newcommand{\tnote}[1]{\par\vspace{2pt}\noindent\begin{minipage}{\linewidth}\footnotesize #1\end{minipage}}

\begin{document}

\title{DEMON: Diffusion Engine for Musical Orchestrated Noise\\[2pt]
{\large High-throughput streaming diffusion as a real-time musical control surface}}

\author{\IEEEauthorblockN{Ryan Fosdick}
\IEEEauthorblockA{Daydream\\[3pt]
\href{mailto:ryan@livepeer.org}{ryan@livepeer.org}\\[1pt]
Project page (audio, demo video): \url{https://daydreamlive.github.io/DEMON/}\\[1pt]
Code: \url{https://github.com/daydreamlive/DEMON}}}

\maketitle

\begin{abstract}
We present DEMON, a real-time diffusion engine that makes the denoising process playable as a live musical instrument: a control surface both broad (many parameters shaped per-frame across the output) and responsive (each control taking effect as fast as its place in the denoising loop allows). Built on ACE-Step 1.5 and StreamDiffusion's ring-buffer architecture with TensorRT acceleration, it sustains up to 12.3 decoder completions per second for 60-second music on a single consumer GPU (RTX 5090), or 11.3 generations per second at our production ring-depth of 4. At these rates denoising parameters become viable as live performance controls, but the ring buffer propagates per-request changes only at its drain rate, a floor of S denoising steps. We contribute four mechanisms. (1) Per-slot heterogeneous denoise scheduling: each ring-buffer slot owns its timestep schedule, so a moving denoise slider is tracked without wiping the in-flight queue, where the upstream global-schedule design must rebuild and discard it. (2) Shared mutable per-step state, giving any parameter consulted at every solver step next-tick effect, bypassing ring-buffer drain. (3) Per-frame source blending: a sampling-time control on the standard SDE re-noise step, giving a framewise transformation-strength axis that complements scalar denoise scheduling. (4) Windowed VAE decode exploiting receptive-field analysis for an 8.0x decode speedup. Together these separate streaming-diffusion parameters into four propagation classes, by onset and convergence latency.
\end{abstract}

\section{Introduction}\label{sec:intro}

Music generation models have reached impressive quality, but most remain batch-mode tools: the user specifies a prompt, waits, and listens. Recent autoregressive systems such as Lyria RealTime \cite{b20} have achieved streaming output with interactive control, generating 2-second audio chunks with adjustable style, tempo, and density. However, in existing chunked autoregressive streaming systems, control updates are typically applied only at chunk boundaries, introducing a practical temporal granularity limit. Because tokens are generated and committed sequentially, sub-chunk control modification generally requires re-decoding, rollback, or specialized architectural mechanisms. For musicians accustomed to the continuous, sub-millisecond response of synthesizers and DAWs, chunk-boundary control latency remains coarse.

Diffusion models \cite{b12} offer an alternative path. StreamDiffusion \cite{b1} demonstrated that a ring buffer of in-flight generations at staggered denoising stages achieves one output per forward pass, bringing image diffusion to interactive rates. StreamV2V \cite{b2} extended this to video-to-video translation. We adapt this architecture to ACE-Step 1.5 \cite{b3}, a DiT-based flow-matching music generation model, and combine it with TensorRT acceleration and a windowed VAE decode strategy to achieve 12.3 generations per second for 60-second music on a single consumer GPU (RTX 5090), with the same engine validated out to 240s.

This throughput creates an opportunity and a problem simultaneously. The opportunity: at 12+ generations per second, denoising parameters become viable as live performance controls, updated faster than a musician can perceive, and because the latent is temporal many can also be shaped per-frame. The aim is a playable musical instrument: a control surface both broad in what it exposes and fast enough to perform with. The problem: the ring buffer that enables this throughput also constrains parameter-update responsiveness. Per-request parameters are baked into each slot at submission time and propagate only when a new-conditioning slot completes its full denoising schedule, requiring approximately S ticks (the schedule length) for the first effect to be visible. At depth 8 with S=8, this means $\sim$650ms before any per-request parameter change takes effect; at depth 4, our production operating point, the measured first-effect falls to $\sim$470ms for an $\sim$8\% throughput cost (Section~\ref{sec:depth}).

Adapting the streaming pipeline to long-form audio also required solving domain-specific engineering challenges: a VAE decode bottleneck ($\sim$56ms, nearly matching the DiT forward pass) and a TensorRT precision issue where fp16 quantization compounds through 24 DiT layers. We present DEMON, built on ACE-Step 1.5. Our contributions:

\begin{enumerate}
\item \textbf{Per-slot heterogeneous denoise scheduling.} Each ring-buffer slot owns its own timestep schedule, baked from the denoise value at admission, rather than drawing timesteps from a single global tensor (the upstream StreamDiffusion design, whose strength changes go through a queue-resetting \texttt{prepare()} rebuild that discards all in-flight state). The batched forward draws each row's timestep from its own slot, so a moving denoise slider is tracked across submissions without a queue wipe and every finished output is a coherent single-schedule trajectory; a change still reaches the output at the S-tick drain floor, but no in-flight work is discarded and no warmup is repaid. Against a StreamDiffusion-style global-reset baseline under a continuous slider sweep, this sustains a 100\% completion rate where the baseline produces output on 1.7\% of ticks (Sections~\ref{sec:pipeline}, \ref{sec:ablation}).

\item \textbf{Shared mutable per-step state} that bypasses ring-buffer drain. Because per-step parameters are consulted at every denoising step rather than once at submission, they can be held as pipeline-level mutable state rather than frozen per-slot state; when updated, all in-flight slots use the new value on the very next tick, with effect proportional to remaining steps. Its instances are the per-frame curves enumerated in Section~\ref{sec:engine} (the SDE source-blending curve of contribution 3 among them), with the x0-target morph as the worked example (Section~\ref{sec:pipeline}). The same pattern applied to the structural timestep schedule yields \emph{in-flight schedule migration}, a fast-onset option for denoise we treat as a special case (Section~\ref{sec:pipeline}). Together these populate a four-class taxonomy of parameter propagation (per-request, migrated-schedule, per-step shared-mutable, model-weight; Section~\ref{sec:pipeline}), applicable to any ring-buffer architecture including StreamDiffusion for images.

\item \textbf{Per-frame source blending on the SDE re-noise step,} a sampling-time control in which a shaped curve sets, independently at each temporal frame, how far each step is re-noised toward the model prediction vs.\ anchored to the source latents. It produces a source-preservation gradient that the global scalar denoise does not provide (consistent across 66 samples, Section~\ref{sec:perframe}), and complements the per-slot heterogeneous scheduling of contribution 1 (Section~\ref{sec:engine}).

\item \textbf{Windowed VAE decode} that exploits empirical receptive-field analysis of the Oobleck VAE to decode only the playback window with overlap margins, achieving sample-identical interior output (16-bit PCM render) at 8.0x lower latency (Section~\ref{sec:vae}).
\end{enumerate}

Our pipeline builds on three elements of StreamDiffusion \cite{b1}, none claimed as contributions here: we adapt its ring-buffer streaming architecture to variable-length audio latents (Section~\ref{sec:pipeline}), the substrate on which the per-slot scheduling of contribution 1 is built; we implement its residual CFG (RCFG) in both onetime-negative and self-negative variants to cut the per-step unconditional forward pass (Section~\ref{sec:multicond}); and we apply the \emph{idea} of its stochastic similarity filter as an inference-time decode skip (Section~\ref{sec:filter}). Our similarity filter is a deterministic latent-MSE threshold gating the VAE decode, not the original's probabilistic, cosine-similarity skip of the denoising pass, so we credit the concept rather than claim a faithful reimplementation.

\section{Related Work}\label{sec:related}

\subsection{Music Generation Models}\label{sec:related-models}

Recent music generation spans autoregressive and diffusion approaches. MusicGen \cite{b6} uses a single-stage autoregressive transformer over discrete audio codes with text and melody conditioning; at 3.3B parameters, it generates slower than real-time (41.3s for 30s of audio on an A100 \cite{b21}). MAGNeT \cite{b21} achieves a 7x speedup over MusicGen via non-autoregressive masked parallel decoding, but generates fixed-length clips without real-time streaming or fine-grained interactive temporal control. AudioLDM \cite{b8} applies latent diffusion \cite{b11} to audio spectrograms. Stable Audio \cite{b7} introduces timing conditioning for variable-length latent diffusion. Presto \cite{b22} distills both steps and layers from a diffusion model, achieving remarkable batch throughput (230ms for 32s of audio on A100), but does not target real-time streaming or interactive generation. ACE-Step \cite{b3} adopts a Diffusion Transformer (DiT) \cite{b5} with flow matching \cite{b4}, operating in a learned latent space at 25Hz with 64-dimensional tokens, with text, timbre-reference, and source-structure conditioning. Its turbo variant achieves generation in 8 denoising steps via few-step distillation.

\subsection{Real-Time Music Generation}\label{sec:related-rtmusic}

Lyria RealTime \cite{b20} (Google DeepMind, 2025) is the most directly comparable system. It is the API-tier model of the ``Live Music Models'' family, whose open-weights sibling, Magenta RealTime, is a 750M-parameter autoregressive transformer that streams 2-second audio chunks; the paper states that Lyria RealTime shares this same general architecture with additional controls, and does not disclose its parameter count. Its interactive controls include tempo, brightness, density, musical key, and per-stem (instrument) toggles. Lyria's chunked autoregressive design imposes practical constraints: control updates take effect only at chunk boundaries (minimum 2-second latency), there is no per-frame modulation within a chunk, and committed tokens are not revised. We compare quantitatively in Table~\ref{tab:syscompare}. Real-time neural audio synthesis systems such as DDSP \cite{b18} and RAVE \cite{b19} achieve interactive rates with lightweight architectures but operate at the waveform level rather than the semantic level of a full diffusion model.

\subsection{Streaming Diffusion}\label{sec:related-streaming}

StreamDiffusion \cite{b1} introduced the key insight that a ring buffer (Stream Batch) of in-flight generations at staggered denoising stages can achieve one-output-per-forward-pass throughput after a warmup period, bringing image diffusion to interactive rates. StreamV2V \cite{b2} extended this to video-to-video translation by maintaining a feature bank of past-frame features and fusing them into the current frame through extended self-attention. Latent consistency models (LCM) \cite{b14} reduce the number of required denoising steps, complementing streaming approaches. Our work adapts StreamDiffusion's ring-buffer architecture to the audio domain, where the temporal extent of each generation (60--240s, 1500--6000 frames at 25Hz) and the cost of VAE decoding introduce challenges absent in image streaming.

\subsection{Controllable Diffusion}\label{sec:related-control}

ControlNet \cite{b15} and T2I-Adapter \cite{b16} demonstrated that auxiliary conditioning can steer diffusion generation without retraining the base model, primarily for spatial control in images. IP-Adapter \cite{b17} extended this to image-prompt conditioning. In the audio domain, control has primarily been achieved through text prompts, timing metadata, melody conditioning, and audio-reference conditioning, the last including global timbre/style reference and per-request source-structure conditioning (e.g.\ ACE-Step \cite{b3}). Dynamic per-step control of the denoising process is by now well developed for images (e.g.\ dynamic CFG), and per-frame modulation of denoising parameters has been applied to music offline \cite{b26}; doing so approaching real time, within a streaming pipeline, has not. DEMON exploits this gap (Section~\ref{sec:engine}).

\section{Architecture}\label{sec:arch}

\subsection{System Overview}\label{sec:overview}

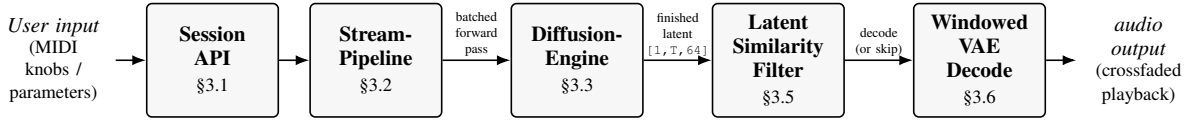
\begin{figure*}[!t]
\centering
\begin{tikzpicture}[
  font=\footnotesize,
  >={Latex[length=1.8mm]},
  proc/.style={draw, semithick, rounded corners=2pt, align=center,
               text width=16mm, minimum height=15mm, inner sep=2pt, fill=black!3},
  term/.style={align=center, text width=14mm, font=\footnotesize},
  lbl/.style={midway, above, font=\tiny, align=center, inner sep=1.5pt},
  flow/.style={->, semithick},
]
\node[term] (in) {\textit{User input}\\{\scriptsize(MIDI knobs /\\parameters)}};
\node[proc, right=4mm of in]     (sess)   {\textbf{Session\\API}\\[1pt]{\scriptsize\S3.1}};
\node[proc, right=4mm of sess]   (stream) {\textbf{Stream-\\Pipeline}\\[1pt]{\scriptsize\S3.2}};
\node[proc, right=9mm of stream] (engine) {\textbf{Diffusion-\\Engine}\\[1pt]{\scriptsize\S3.3}};
\node[proc, right=9mm of engine] (filt)   {\textbf{Latent\\Similarity\\Filter}\\[1pt]{\scriptsize\S3.5}};
\node[proc, right=9mm of filt]   (vae)    {\textbf{Windowed\\VAE\\Decode}\\[1pt]{\scriptsize\S3.6}};
\node[term, right=4mm of vae]    (out)    {\textit{audio output}\\{\scriptsize(crossfaded\\playback)}};
\draw[flow] (in) -- (sess);
\draw[flow] (sess) -- (stream);
\draw[flow] (stream) -- node[lbl]{batched\\forward\\pass} (engine);
\draw[flow] (engine) -- node[lbl]{finished\\latent\\\texttt{[1,T,64]}} (filt);
\draw[flow] (filt) -- node[lbl]{decode\\(or skip)} (vae);
\draw[flow] (vae) -- (out);
\end{tikzpicture}
\caption{System architecture. User input (MIDI knobs / parameters) drives the Session API (text encoding, source preparation, LoRA management, caching), which feeds the StreamPipeline: a ring buffer of 8 in-flight generations at staggered denoising stages (\texttt{submit()}\,$\rightarrow$\,\texttt{tick()}\,$\rightarrow$\,latent). Each tick runs one batched forward pass through the DiffusionEngine (SDE source blending, per-frame curves, multi-condition; ODE/SDE solver). The finished latent passes a latent similarity filter (skip the VAE decode and reuse the last audio when MSE against the previous latent is below threshold) before windowed VAE decode (only the playback window, with overlap margins and sample-identical interior); the resulting chunk is crossfaded into the output stream.}
\label{fig:arch}
\end{figure*}

DEMON is organized as a five-stage pipeline from user input to audio output (Figure~\ref{fig:arch}). The \emph{Session API} is the front door: it encodes text, prepares source audio, manages LoRAs, and caches model loading and torch.compile warmup across generations. The \emph{StreamPipeline} (Section~\ref{sec:pipeline}) is the streaming core, a ring buffer of in-flight generations held at staggered denoising stages that emits one finished latent per \texttt{tick()}; it is where the two streaming-specific control mechanisms of this paper live (per-slot heterogeneous scheduling and shared mutable per-step state). Each tick runs one batched forward pass through the \emph{Diffusion Engine} (Section~\ref{sec:engine}), which owns the ODE/SDE solver and the per-frame control curves that shape each step. The finished latent then passes the \emph{Latent Similarity Filter} (Section~\ref{sec:filter}), which reuses the previous decode when the latent is nearly unchanged, before \emph{Windowed VAE Decode} (Section~\ref{sec:vae}) renders only the playback window to audio. Cutting across every stage, an \emph{Acceleration} layer (Section~\ref{sec:accel}: TensorRT mixed-precision engines and runtime LoRA refit) brings the system to real time. We describe each below, devoting the most detail to the StreamPipeline, where the streaming-specific contributions are concentrated.

\subsection{Streaming Pipeline}\label{sec:pipeline}

The StreamPipeline maintains a ring buffer of in-flight generations at different denoising stages, adapted from StreamDiffusion \cite{b1}. Each \texttt{tick()} performs one batched forward pass advancing all slots simultaneously. After warmup, at depth D with S denoising steps, one completion emerges every S/D ticks.

Critically, the ring buffer depth is decoupled from the denoising step count. Both are tunable; depth D is a property of the streaming pipeline, not of the base model, and controls only how many generations are in flight simultaneously (and therefore the batch size of each forward pass), while the step count S is the per-generation denoising trajectory, for which our experiments use ACE-Step turbo's few-step default of 8. This decoupling enables a throughput-vs-responsiveness tradeoff with no quality difference: higher depth increases batch efficiency (more gens/sec) at the cost of slower parameter-change propagation. We characterize this tradeoff in Section~\ref{sec:propagation}.

Our audio-domain adaptations over the image-domain StreamDiffusion: we make slots first-class state-carrying objects (with their own timestep schedule, SDE curve snapshot, and source latents) rather than positional slots in a shared queue. Per-slot curve and source state lets the streaming pipeline support the per-frame modulation system of Section~\ref{sec:engine} without all in-flight generations sharing the same modulation. Two mechanisms exploit this per-slot state to keep denoising parameters responsive as live controls: per-slot heterogeneous denoise scheduling, and shared mutable per-step state.

\textbf{Per-slot heterogeneous denoise scheduling.} StreamDiffusion's ring buffer is governed by a single global timestep tensor stored on the pipeline. In its \texttt{prepare()} path, a subset of the scheduler's timesteps is selected once at session setup and broadcast across the buffer; in the batched forward, all rows draw timesteps from this shared tensor by position (row k always carries the k-th denoising step's timestep). Changing strength requires calling \texttt{prepare()} again, which rebuilds the timestep tensor and zeros out the rolling latent buffer. The in-flight queue is wiped.

A later derivative, the Daydream StreamDiffusion fork \cite{b23} (an earlier implementation of ours), relaxes this: it adds a runtime updater that edits the global timestep list in place, so a same-length value change (a strength adjustment) updates the schedule without zeroing the buffer, while a step-count change still triggers a full rebuild. That schedule is nonetheless a single global list shared by every in-flight frame: a strength change switches all of them together on the same tick, and the design still cannot hold different frames at different strengths within one forward pass.

For music streaming over 60-second outputs, the picture is different. Denoise is the most musical control on the system (Section~\ref{sec:paramlatency}) and a natural target for continuous adjustment during a session. Wiping the queue on every slider movement would produce audible dead-air gaps as the buffer refills, defeating the purpose of the streaming pipeline.

We resolve this by making slots first-class objects with their own scheduling state. Each \texttt{\_Slot} carries its own timestep schedule, baked at admission from the denoise value at that moment, plus a per-slot step counter. The pipeline maintains a small cache keyed by denoise value, so repeated submissions at the same strength share a schedule tensor and pay no allocation cost. The batched forward pass at each tick draws each row's timestep from its own slot's schedule, \texttt{[slot.t\_schedule[slot.step] for slot in active\_slots]}, so the DiT receives a heterogeneous timestep tensor in which different rows are at different denoise strengths simultaneously.

The user-facing consequence is that the queue never has to be drained-and-refilled. When the denoise slider moves between submissions, the next admitted slot uses the new schedule, in-flight slots continue to their natural completion on the schedules they were born with, and the output stream continues without interruption. Every finished output is therefore a coherent single-schedule trajectory. This does not, however, reduce onset latency: the time until a new denoise value is reflected in the output remains the natural S-tick floor of the ring buffer (one new-conditioning slot completing its schedule), even though no in-flight work is discarded and no buffer warmup is paid. The closest prior point we are aware of is the global mutable schedule above \cite{b23}, which makes the timestep list editable at runtime but per pipeline rather than per slot. We are not aware of streaming-diffusion work, including for audio, that makes the timestep schedule first-class \emph{per-slot} state, the property that lets a single batched forward carry rows at different denoise strengths at once.

Section~\ref{sec:ablation} substantiates the throughput-preservation claim with a direct ablation against a StreamDiffusion-style global-reset baseline. This is our primary denoise-control mechanism; the S-tick onset floor it leaves in place can optionally be traded for a faster onset via in-flight schedule migration, characterized as a special case of shared-mutable state below.

\textbf{Shared mutable per-step state.} Per-slot scheduling above preserves throughput across a moving denoise slider, but at the S-tick onset floor. A different mechanism underlies the system's faster-than-drain controls: a parameter consulted at every denoising step, rather than once at submission, can be held as \emph{shared mutable state}, improving onset (and, for per-step parameters, convergence) relative to the per-request floor. In-flight schedule migration (defined below) is the one application of the pattern to a structural rather than per-step parameter.

That convergence floor is approximately S$\cdot$tick\_ms (where tick\_ms grows with depth due to batching cost; see Section~\ref{sec:depth} for the depth--latency curve). LoRA refit bypasses it by modifying shared model weights (Section~\ref{sec:accel}). The same principle applies to any denoising parameter consulted per-step: the per-frame curves enumerated in Section~\ref{sec:engine} are all read at every step, so each can be held as shared state on the pipeline rather than frozen per-slot state, settable for all in-flight slots through a single field-name-keyed registry.

When a shared curve is set, all in-flight slots read it on the very next tick, regardless of pipeline depth. The effect on each slot is proportional to its remaining denoising steps: a slot at step 0 (all steps remaining) experiences the full effect of the new curve, while a slot at step 7 (one step remaining) is minimally affected. This produces progressive convergence across completions rather than the step-function drain of per-request parameters.

These mechanisms are complementary, not redundant, and they separate along two latency axes: \emph{onset} (when the change first reaches the output) and \emph{convergence} (when the output fully reflects it). The per-slot schedule above governs the parameter that defines a slot's whole trajectory; migrating it (defined below) achieves 1-tick onset but, because migration produces trajectory-incoherent hybrids, retains the S-tick convergence floor. Per-step shared-mutable parameters are read fresh at every step, so a change onsets in one tick \emph{and} converges progressively below the S-tick floor (their partial applications are convergent, not hybrids). Model-weight changes (LoRA refit) are immediate on both axes. Together they identify four propagation classes in streaming diffusion pipelines (Table~\ref{tab:classes}):

\begin{table*}[!t]
\caption{Four propagation classes in streaming-diffusion pipelines.}\label{tab:classes}
\centering
\begin{tabular}{@{}p{0.155\textwidth}p{0.165\textwidth}p{0.065\textwidth}p{0.135\textwidth}p{0.30\textwidth}@{}}
\toprule
Class & Examples & Onset & Convergence & Mechanism \\
\midrule
Per-request (frozen) & Conditioning, source audio & S ticks & S ticks & Baked into slot at submission \\
\addlinespace
Schedule, migrated (shared-mutable) & Denoise schedule (migration) & 1 tick & S ticks & Migrated onto every in-flight slot; transient hybrids \\
\addlinespace
Per-step shared-mutable (shadowed) & SDE curve, x0-target morph, velocity scale & 1 tick & Progressive ($< $ S) & Read from shared state each step \\
\addlinespace
Model weights & LoRA refit & 1 tick & 1 tick & Shared decoder weights \\
\bottomrule
\end{tabular}
\tnote{The migrated-schedule and per-step shared-mutable rows both read ``1 tick'' onset, but it differs in kind: a per-step parameter's early completions are convergent partial applications, while the migrated schedule's are trajectory-incoherent hybrids that change the output without yet approaching the new-denoise target (defined below). The ``1 tick'' entry marks when the output first changes, not when it first moves toward the requested value.}
\end{table*}

\textbf{In-flight schedule migration} treats the per-slot timestep schedule as shared mutable state: a pass at the top of each tick reassigns every in-flight slot to the new schedule, holding its step index fixed and swapping only the sigma values (the schedule always has $S{+}1$ entries regardless of denoise, so a slot $k$ steps in stays $k$ steps in). This drops a denoise change's onset from $S$ ticks to one, but is deliberately not trajectory-coherent: a latent integrated to the old schedule's sigma is reinterpreted at the new one, so the swap-tick completions are hybrids that depart from the previous output without approaching the new-denoise target until a slot has run all $S$ steps on the new schedule. Migration therefore occupies a distinct (1-tick onset, $S$-tick convergence) regime, and the benefit appears only at depth $\geq 2$ (largest at the depth-8 floor: first-effect tick~1 vs.\ tick~8). Mechanically this is the in-place global-schedule edit of the daydreamlive StreamDiffusion fork \cite{b23} expressed in per-slot state, so we credit that earlier form and present migration as a characterized fast-onset option rather than a contribution. On our own listening the faster onset carries a transient muffling until the buffer drains (Section~\ref{sec:discussion}), so heterogeneous scheduling above, not migration, is our denoise control surface.

\textbf{Worked example: x0-target morph.} \texttt{x0\_target\_strength} blends the per-step $x_0$ prediction toward a precomputed target latent, $x_0 \leftarrow (1-\alpha)\,x_0 + \alpha\,x_0^{\text{target}}$, per-frame and gated to the refinement half of the schedule. Because $\alpha$ is read through the shared registry every step, a live write takes effect on the next denoising pass and converges over a four-step ramp (set by the refinement-half gate): the per-step class's 1-tick-onset, sub-$S$-convergence signature. The same field carries both control axes: a shaped per-frame $\alpha$ curve produces a progressive morph across the song (the spatial axis of Section~\ref{sec:engine}), while a live \texttt{set\_shared\_curve} write produces a tick-granularity onset (the onset axis above). Unlike the SDE source blending of Section~\ref{sec:engine}, which can only anchor toward the \emph{source}, the target here is an independent latent: the canonical workflow covers a track, banks several cover variants, and blends toward them at tick granularity. Section~\ref{sec:perframe} measures the morph gradient; Table~\ref{tab:propagation} reports its live onset.

\subsection{Diffusion Engine}\label{sec:engine}

The Diffusion Engine runs the per-step denoising math each tick: the ODE/SDE solver and the per-frame curves that shape it. These per-frame curves are the breadth axis of the control surface (Section~\ref{sec:intro}): the many parameters the instrument exposes, shaped at 25~Hz, complementing the responsiveness mechanisms of Section~\ref{sec:pipeline}. Its defining control is a source-blending step on the SDE re-noise; the full set of per-frame curves it reads from shared state follows.

\textbf{SDE with per-frame source blending.} The \texttt{denoise} parameter is the single most impactful generation control: it determines transformation strength by truncating the timestep schedule. The Streaming Pipeline (Section~\ref{sec:pipeline}) addressed denoise on the \emph{temporal} axis, how fast a change to the scalar reaches the output without wiping the queue (and how schedule migration there trades trajectory coherence for a faster onset). This control addresses a different axis: \emph{spatial} resolution, whether transformation strength can vary across the 1500 frames of a single generation. Scalar denoise cannot, because truncating the schedule applies globally to every frame alike. We expose framewise transformation-strength control through a source-blending control on the SDE re-noise step, a complement to that per-slot heterogeneous scheduling.

Standard diffusion uses either an ODE solver (deterministic) or an SDE solver (stochastic re-noising at each step) \cite{b13}. We keep the standard SDE step unchanged and add a per-frame control on top of it, decomposing ``how much to transform'' into a framewise operation.

At each denoising step, the solver computes two re-noised candidates and blends between them per-frame:

\begin{figure*}[!t]
\textbf{SDE re-noise step with per-frame source blending.}
\begin{lstlisting}[style=pseudo]
x0_pred = xt - vt * t_curr                                    # velocity to x0
sde_noise = randn_like(xt)                                     # fresh noise
xt_full   = t_next * sde_noise + (1 - t_next) * x0_pred       # standard SDE re-noise
xt_source = t_next * sde_noise + (1 - t_next) * source_latents # source-anchored re-noise
xt_next   = curve[t] * xt_full + (1 - curve[t]) * xt_source   # per-frame blend
\end{lstlisting}
\end{figure*}

When \texttt{curve=1.0}, this reduces to standard SDE (full re-noise from the model's x0 prediction). When \texttt{curve=0.0}, the re-noised state is anchored to the source latents, pulling the output toward the original audio at each step. Values in between provide continuous blending. Because the curve operates per-frame, a shaped curve (ramp from 0 to 1 over 60 seconds) makes different temporal regions of the output behave differently within a single generation: the beginning preserves the source while the end is fully generated.

This control is distinct from the ODE denoise parameter in two ways. First, denoise truncates the timestep schedule globally (fewer steps for every frame alike), so it cannot express a per-frame gradient; the SDE curve can. Second, the SDE curve modulates behavior at every step for every frame, a finer granularity of control. The two act on different axes: per-slot heterogeneous scheduling (Section~\ref{sec:pipeline}) keeps the \emph{scalar} denoise tracking a moving slider, while the SDE curve adds \emph{framewise} transformation-strength control. Its mechanism (source-anchored re-noise blending) and acoustic signature differ from schedule truncation, so it is its own control axis, not an approximation of denoise.

\textbf{Per-frame denoising dynamics.} The streaming architecture operates on latent tensors of shape \texttt{[B, T, D]} where T = 1500 frames at 25Hz. A denoising parameter that acts element-wise along this temporal dimension can therefore be generalized from a scalar to a per-frame curve with no architectural modification, though whether the result is stable and musically meaningful is parameter-specific; the clearest exception is denoise itself, which truncates the schedule globally and so cannot be made per-frame at all (the source-blending control above operates differently). The curve is a tensor of shape \texttt{[T]} or \texttt{[B,T]}, broadcast to \texttt{[B,T,1]} for element-wise operations. When no curve is provided, a sentinel value (all-ones for multiplicative, all-zeros for additive) is used, making the compiled fast path branch-free.

The engine exposes the per-frame denoising-dynamics curves of Table~\ref{tab:curves}:

\begin{table*}[!t]
\caption{Per-frame denoising-dynamics curves exposed by the engine. Each is a $[1,T,1]$ tensor resolved per-step, so all are hot-mutable as shared state (Section~\ref{sec:pipeline}); the CFG family (\texttt{guidance\_curve}, \texttt{apg\_momentum}, \texttt{cfg\_rescale\_curve}) only modulates a slot whose CFG is already active.}\label{tab:curves}
\centering
\begin{tabular}{@{}lp{0.35\textwidth}p{0.37\textwidth}@{}}
\toprule
Curve & Operation & Effect \\
\midrule
\texttt{sde\_denoise\_curve} & Modulates SDE re-noise blending per-frame (above) & Per-frame control over source preservation vs.\ exploration: a ramp curve yields a monotone per-frame source-preservation gradient (Section~\ref{sec:perframe}). \\
\addlinespace
\texttt{guidance\_curve} & Per-frame CFG scale: \texttt{v = v\_uncond + scale(t) * (v\_cond - v\_uncond)} & Dynamic guidance strength. High at start for structural adherence, low at end for natural decay. \\
\addlinespace
\texttt{velocity\_scale} & Per-frame multiplier on the model velocity $v_t$ before integration & Scales the transformation rate per frame: $<1$ damps motion toward $x_0$, $>1$ accelerates it. \\
\addlinespace
\texttt{ode\_noise\_curve} & Per-frame stochastic noise added after the deterministic ODE step & Injects controllable per-frame texture/variation on the otherwise deterministic path. \\
\addlinespace
\texttt{apg\_momentum} & Per-frame momentum coefficient for APG guidance & Shapes the guidance momentum buffer across the timeline. \\
\addlinespace
\texttt{cfg\_rescale\_curve} & Per-frame blend of the APG output norm back toward the positive-velocity norm & Per-frame guidance-magnitude rescaling. \\
\addlinespace
\texttt{x0\_target\_strength} & Per-frame blend of the $x_0$ prediction toward a precomputed target latent, gated to the refinement half (Section~\ref{sec:pipeline}) & Per-frame morph toward an independent target (e.g.\ a banked cover variant); shaped curves morph progressively across the song. \\
\bottomrule
\end{tabular}
\end{table*}

\subsection{Multi-Condition Support}\label{sec:multicond}

The engine supports composable multi-condition generation: velocity fields from active conditions are blended by per-frame temporal weights (\texttt{v\_blended = sum(w\_i v\_i) / sum(w\_i)}), enabling prompt crossfades, negative conditioning (CFG), and multi-LoRA composition, with automatic routing to the most efficient path by condition count and LoRA identity. We also implement StreamDiffusion's residual CFG (RCFG) \cite{b1} in onetime-negative and self-negative variants, removing the per-step unconditional forward pass that standard CFG requires.

\subsection{Latent Similarity Filter}\label{sec:filter}

A second decode optimization sits between the engine and the VAE: when consecutive completions produce nearly identical latents (\texttt{MSE(latent\_curr, latent\_prev)} below 1e-3), we skip the VAE call and reuse the previous decoded audio. This adapts the \emph{idea} of StreamDiffusion's stochastic similarity filter \cite{b1} to audio: our gate is a deterministic latent-MSE threshold on the VAE decode, not the original's probabilistic cosine-similarity skip of the denoising pass. When per-request parameters are held steady, successive completions are deterministic re-renderings of the same conditioning, so the skip triggers often; on the cross-GPU cover benchmark the skip rate scales with depth (9--18\% at depth 1, 45--53\% at depth 4, 56--65\% at depth 8), since more in-flight slots share conditioning between completions. The amortized decode cost drops during stable regions with no perceptible quality change.

\subsection{Windowed VAE Decode}\label{sec:vae}

ACE-Step generates variable-length latent at 25Hz ($\sim$1500 frames at our 60-second operating point), but in streaming mode the user only hears a few seconds at a time. Full VAE decode of the entire latent is the second-largest latency component ($\sim$56ms with TRT). Windowed decode addresses this by processing only the playback window.

The Oobleck VAE uses dilated 1D convolutions with receptive fields spanning multiple seconds. Naively slicing the latent produces boundary artifacts. Through empirical measurement, we find that the receptive field converges within 333ms (8.3 frames) of the boundary: beyond this distance, the windowed interior matches the full decode sample-for-sample at the 16-bit PCM render (Section~\ref{sec:vaequality}).

The windowed decode procedure:
\begin{enumerate}
\item Extend the target window by overlap margins on both sides (default 0.5s, providing 12.5 frames of context beyond the convergence threshold).
\item Decode the extended region through the VAE.
\item Trim the overlap margins, returning only the clean interior with sample-accurate alignment to 25fps frame boundaries (1920 samples/frame at 48kHz).
\end{enumerate}

The speedup is a function of the chosen playback window. With a 3s window and 0.5s overlap, VAE decode drops from 56ms to 7ms (8.0x); with a 15s window it drops to 20ms (2.8x), the operating point used in the cross-GPU benchmark of Section~\ref{sec:crossgpu}. The interior matches the full decode sample-for-sample (16-bit PCM) at every window size; the smaller the window, the larger the saving.

Because the window is fixed by the playback need rather than the generation length, windowed VAE decode cost is independent of song duration. The DiT decoder forward pass is then the only latency component that grows with generation length (Section~\ref{sec:e2e}). This is what makes generation duration a clean operating-point choice rather than a fixed system property: we run 60s as the primary operating point and validate a 240s profile (Sections~\ref{sec:setup}, \ref{sec:complatency}), with the same streaming, scheduling, and VAE machinery unchanged.

\subsection{Acceleration}\label{sec:accel}

Several layers of acceleration bring the system to real-time performance:

\textbf{TensorRT mixed-precision engines.} The DiT decoder is exported to ONNX with trace-safe patching (replacing un-traceable Lambda modules, disabling GQA for SDPA traceability, converting shape-dependent Python branches to tensor operations). The ONNX graph uses a mixed-precision strategy: attention and MLP blocks in fp16, while timestep embeddings, AdaLN scale-shift tables, and RMSNorm layers remain in fp32. This prevents accumulative gate error through 24 DiT layers. Empirical comparison of all-fp16 and mixed-precision engines on identical inputs shows the all-fp16 engine attenuates the velocity output by $\sim$7x (output magnitude ratio 0.13--0.16x across timesteps, consistent with $\sim$0.92x compounding per layer: $0.92^{24} \approx 0.14$), and produces NaN at early timesteps ($t < 0.5$). The TRT engine is built with the STRONGLY\_TYPED flag to preserve these precision annotations.

\textbf{Runtime LoRA refit.} TensorRT engines are typically static, requiring a full rebuild to change weights. We enable the REFIT builder flag and export ONNX with constant folding disabled (preserving PyTorch parameter names as ONNX initializer names). At runtime, the TRT IRefitter API allows us to compute LoRA deltas (B @ A in fp32, convert to engine dtype) and apply them to the running engine without rebuild. Multiple LoRAs stack additively with independent strengths, and refit-enabled engines show no measurable latency penalty at base weights (Table~\ref{tab:complatency}).

\textbf{VAE TRT engines.} Separate encode and decode engines with dynamic shape profiles support the windowed decode workflow. A shared Polygraphy CUDA stream across all TRT engines avoids stream synchronization overhead.

\subsection{Composability}\label{sec:composability}

These control dimensions (per-frame curves, conditioning composition, source manipulation, LoRA, solver choice) compose freely because each acts at a different point in the execution path (curves and solver inside the per-step DiT call, conditioning at velocity-blend time, source at noise construction and SDE re-noise, LoRA on the decoder weights), so a session can layer several at once with no per-workflow plumbing. Our reference workflows span this space, from a prompt-only cover to a production session combining curves, dual-prompt CFG, masked source, and stacked LoRAs, each a self-contained script with zero engine modification.

\section{Experiments}\label{sec:experiments}

Our experiments fall into four groups: latency, from per-component to end-to-end (Sections~\ref{sec:syscompare}--\ref{sec:e2e}); output quality, covering the windowed VAE, the SDE and x0-target per-frame control, and stream-vs-batch parity (Sections~\ref{sec:vaequality}--\ref{sec:quality}); parameter-change propagation (Section~\ref{sec:propagation}); and cross-GPU replication (Section~\ref{sec:crossgpu}). The propagation results are the paper's central empirical result, the responsiveness pillar made measurable (the per-frame control results of Section~\ref{sec:perframe} do the same for the breadth pillar). They realize the four-class taxonomy of Section~\ref{sec:pipeline} as deterministic, bit-exact measurements (a per-request change is invisible until tick \emph{k}, then steps to full magnitude; a shared-mutable change takes effect on the next tick), they include the direct heterogeneous-vs-global-reset ablation for our primary contribution (Section~\ref{sec:ablation}), and Section~\ref{sec:crossgpu} shows the same patterns hold within 0.02 RMS across three GPU generations, establishing them as properties of the streaming architecture rather than of one benchmark machine.

\subsection{Setup}\label{sec:setup}

Primary benchmarks run on an NVIDIA RTX 5090 (32GB VRAM; driver 591.86, CUDA 13.1, TensorRT 10.13). Cross-GPU validation on RTX 4090 (24GB) and RTX 3090 (24GB) is reported in Section~\ref{sec:crossgpu}. Model: ACE-Step 1.5 turbo (8 denoising steps, shift=3.0). TRT engines built with 16GB builder workspace, mixed precision (fp16 bulk, fp32 AdaLN/norms), batch\_max=8. Two engine profiles per component: seq\_max=1500 (60s) and seq\_max=6000 (240s), enabling comparison of profile overhead. Audio is 48kHz stereo full-mix (no stem or instrument separation); conditioning is a text prompt plus optional source audio. Quality and control evaluations use six source recordings spanning electronic, ambient, jazz, lo-fi, metal, and progressive rock, each driven with a cross-genre cover prompt; the FAD distributional check (Section~\ref{sec:quality}) compares against a 500-track subset of FMA-small \cite{b24}.

\subsection{System Comparison}\label{sec:syscompare}

Table~\ref{tab:syscompare} compares our system against existing real-time and near-real-time music generation approaches.

\begin{table*}[!t]
\caption{Comparison with real-time and near-real-time music generation systems.}
\label{tab:syscompare}
\centering
\footnotesize
\begin{tabular}{@{}lp{0.115\textwidth}p{0.075\textwidth}p{0.10\textwidth}p{0.255\textwidth}l@{}}
\toprule
System & Architecture & Tick / chunk & Per-frame resolution & Interactive control & Hardware \\
\midrule
\textbf{Ours} & Diffusion (DiT + TRT) & \textbf{81ms (depth=8)} & \textbf{40ms (25Hz per-frame)} & \textbf{Per-frame denoising curves + multi-condition composition} & RTX 5090 (32GB) \\
\addlinespace
Lyria RT \cite{b20} & Autoregressive & 2,000ms & none (per-chunk only) & Semantic controls (tempo, brightness, density, stems, key) & Cloud API \\
\addlinespace
MusicGen-L \cite{b6} & Autoregressive (3.3B) & batch only & N/A & text, melody & A100 \\
MAGNeT \cite{b21} & Masked parallel & batch only & N/A & text & A100 \\
Presto \cite{b22} & Distilled diffusion & batch only & N/A & text & A100 \\
Stable Audio \cite{b7} & Latent diffusion & batch only & N/A & text, timing & A100 \\
\bottomrule
\end{tabular}
\end{table*}

The key distinction is architectural, not incremental: autoregressive streamers such as Lyria commit tokens sequentially, so a control change does not revise emitted audio or modulate within the current chunk (a control floor in current implementations, not a hard theoretical bound), and batch systems expose no mid-generation control at all. DEMON applies its controls inside the per-step denoising loop and along the 1500-frame temporal axis, shaping the denoising dynamics directly at 25~Hz on local consumer hardware. Two axes should not be conflated: per-frame \emph{resolution} (how finely a control is shaped \emph{within} one generation) and control \emph{onset latency} (how quickly a change is \emph{heard}). For Lyria both collapse to the 2-second chunk boundary; DEMON decouples them, with 40~ms (25~Hz) resolution and a configurable onset: one tick ($\sim$43~ms at depth 4) for per-step parameters such as the SDE and guidance curves, and $\sim$470~ms (depth 4) to 649~ms (depth 8) for per-request parameters such as a scalar denoise or prompt change (Section~\ref{sec:depth}). Against the musical-control yardstick, intimate note-level control wants action-to-sound latency near 10~ms \cite{b25}; DEMON's controls are continuous parameter gestures rather than note triggers, and on that looser bar the per-step onset reads as responsive while per-request onset, like Lyria's 2~s chunk, behaves as a discrete command, which is what motivates routing the live controls into the per-step and schedule-migration classes (Section~\ref{sec:pipeline}). The 40~ms figure in Table~\ref{tab:syscompare} is resolution, not onset.

\subsection{Component Latency}\label{sec:complatency}

We profile all TRT engine components across sequence lengths and batch sizes. Two engine profiles are tested per component: 60s (seq\_max=1500) and 240s (seq\_max=6000). At shared operating points the 240s engines match 60s engines within measurement noise ($<$3\%), so the larger profile is strictly more capable with no practical latency cost. Refit-enabled engines (required for runtime LoRA hot-swap) show no overhead at base weights; with a LoRA applied, inference overhead is $\sim$8\% at B=8. LoRA refit itself takes $\sim$1.2s (a one-time cost per style change).

\begin{figure}[!t]
\centering
\includegraphics[width=\columnwidth]{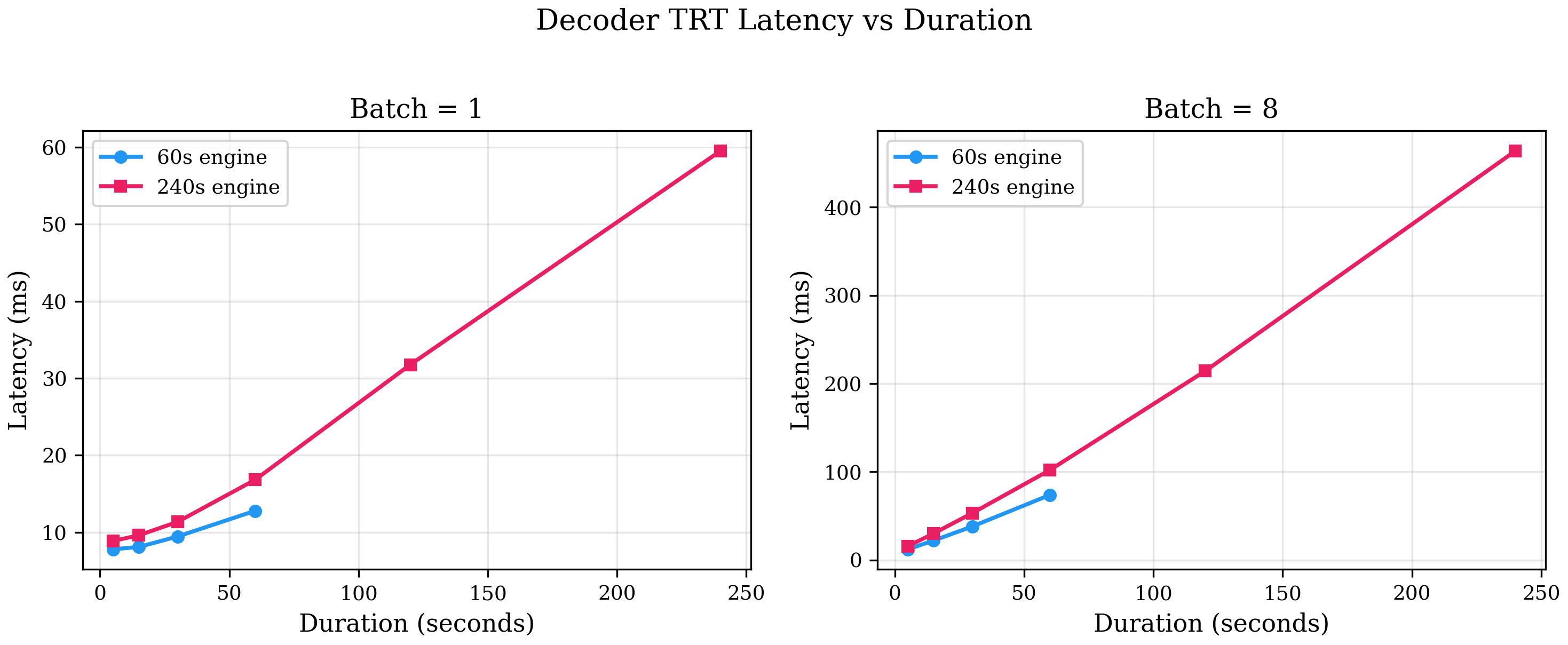}
\caption{Decoder TRT engine forward-pass latency at B=1 and B=8 across sequence lengths. The 60s engine (blue) is rejected by TRT for inputs beyond its 1500-frame profile (the table's ``plateau''); the 240s engine (pink) scales linearly out to 240s. Refit engines only; the table above averages refit and non-refit variants of an earlier build, hence the small absolute offset between figure and table at shared operating points.}
\label{fig:decoder-sweep}
\end{figure}

\begin{table}[!t]
\caption{TRT component forward-pass latency across durations and batch sizes.}
\label{tab:complatency}
\centering
\footnotesize
\setlength{\tabcolsep}{4pt}
\begin{tabular}{@{}llccc@{}}
\toprule
Component & Duration & B=1 (ms) & B=8 (ms) & \begin{tabular}[b]{@{}c@{}}B=8 refit\\+LoRA (ms)\end{tabular} \\
\midrule
\multirow{4}{*}{Decoder} & 5s & 7.0 & 11.4 & 12.6 \\
 & 15s & 7.3 & 23.3 & 26.5 \\
 & 30s & 8.6 & 42.2 & 45.1 \\
 & 60s & 13.3 & 80.6 & 88.1 \\
\midrule
\multirow{4}{*}{VAE decode} & 5s & \multicolumn{2}{c}{10.6} & \\
 & 15s & \multicolumn{2}{c}{18.3} & \\
 & 30s & \multicolumn{2}{c}{30.5} & \\
 & 60s & \multicolumn{2}{c}{55.9} & \\
\midrule
\multirow{4}{*}{VAE encode} & 5s & \multicolumn{2}{c}{12.1} & \\
 & 15s & \multicolumn{2}{c}{19.7} & \\
 & 30s & \multicolumn{2}{c}{32.1} & \\
 & 60s & \multicolumn{2}{c}{56.7} & \\
\bottomrule
\end{tabular}
\tnote{240s-profile engines; 120s and 240s rows omitted (sub-linear scaling continues; see Figure~\ref{fig:decoder-sweep} for the full sweep). Decoder B=8 column refit only. VAE is B=1 only (no batching). Encode is $\sim$1--14\% slower than decode depending on duration (largest gap at the 5s point, smallest at 60s), due to the sampling step on output moments.}
\end{table}

\begin{figure}[!t]
\centering
\includegraphics[width=\columnwidth]{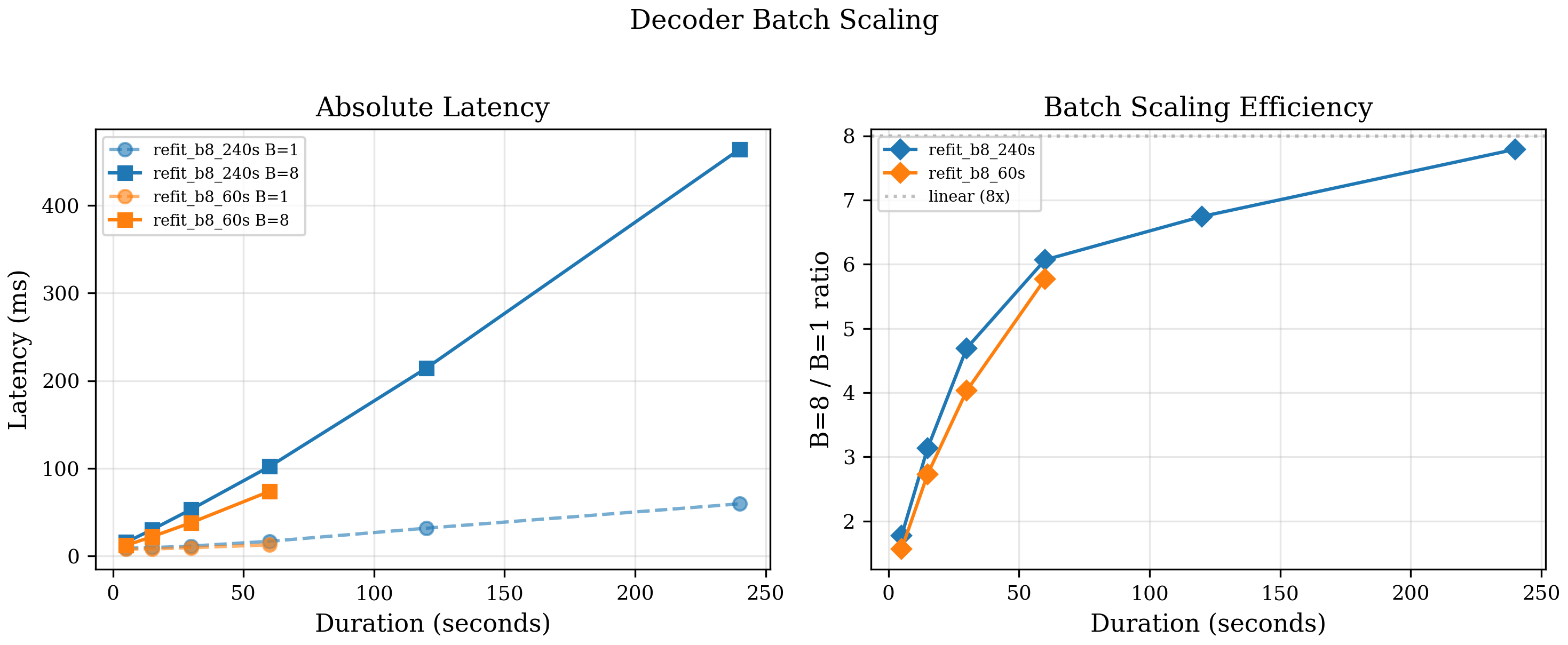}
\caption{Decoder batch scaling. Left: absolute B=1 vs B=8 latency for both engine profiles. Right: B=8/B=1 ratio across durations, sub-linear at short sequences ($\sim$1.6x at 5s) and approaching the 8x linear ceiling at long sequences ($\sim$8.0x at 240s, 240s engine). The intermediate operating points define the throughput-vs-responsiveness tradeoff exploited by pipeline depth.}
\label{fig:batch-scaling}
\end{figure}

Batch scaling is sub-linear (Figure~\ref{fig:batch-scaling}): at the 60s operating point, B=8 costs $\sim$6.1x B=1, well below the 8x linear ceiling. This is what makes higher pipeline depths improve throughput (Section~\ref{sec:propagation}). Intermediate batch sizes lie on the same sub-linear curve; the streaming tick at depth D in Section~\ref{sec:depth} (24.3 ms at D=2, 42.8 ms at D=4) is dominated by the decoder forward pass and serves as a proxy for component cost at B=2 and B=4.

\subsection{End-to-End Streaming Latency}\label{sec:e2e}

Combined decoder tick (B=8, depth=8) plus windowed VAE decode. At 60s generation with a 3s VAE window: 81 + 7 = 88ms per tick. At 240s generation with a 3s window, the decoder scales to 381ms (B=8, measured directly with the 240s engine on the 5090; \texttt{output/decoder\_240s\_5090.json}), giving $\sim$388ms per tick. The VAE window size is independent of generation duration, so windowed decode provides proportionally larger savings at longer durations. Because windowing holds the VAE cost constant across generation length, the DiT decoder forward pass is the sole latency component that scales with song length: the tick grows from 88ms at 60s to $\sim$388ms at 240s entirely through the decoder term. Throughput in generations per second therefore falls with duration (12.3 gens/sec at 60s, $\sim$2.6 at 240s at depth 8), but only because each generation is longer; the audio produced per wall-clock second stays in the same band ($\sim$740s at 60s, $\sim$620s at 240s), hundreds of times faster than real-time at both operating points.

\subsection{Windowed VAE Quality}\label{sec:vaequality}

Windowed decode with 0.5s overlap margins produces interior output that matches the full decode sample-for-sample: the per-sample difference is zero at the 16-bit PCM render, so SNR against the full decode is infinite (noise power 0). This is identity to the rendered-audio precision (a $\geq{\sim}90$ dB floor set by 16-bit quantization), not a float-level claim: differences below the quantization step, including any from TensorRT selecting different convolution kernels for the shorter windowed input, would not register. Without overlap, boundary artifacts appear (29.3 dB SNR, 67\% elevated spectral discontinuity; see Table~\ref{tab:vaequality}).

\begin{table}[!t]
\caption{Windowed VAE decode quality versus full decode.}
\label{tab:vaequality}
\centering
\footnotesize
\begin{tabular}{@{}lcc@{}}
\toprule
Method & SNR vs full & \begin{tabular}[b]{@{}c@{}}Spectral\\discontinuity\end{tabular} \\
\midrule
Full 60s (reference) & ref & 1.000 \\
15s window, 0.5s overlap & identical & 1.706 \\
15s window, no overlap & 29.3 dB & 1.669 \\
5s window, 0.5s overlap & identical & 1.044 \\
\bottomrule
\end{tabular}
\tnote{Spectral discontinuity measures spectral flux at window boundaries relative to the global average (1.0 = no boundary effect). The elevated values for windowed methods reflect hard concatenation at window joins in the evaluation script, not a property of the decode itself: each window's interior matches the corresponding region of the full decode sample-for-sample (16-bit PCM).}
\end{table}

{\footnotesize Windowed-VAE samples are omitted because the table already states sample-identical interior output for the 0.5s-overlap variants. Audio samples for the SDE source-blending and per-curve ablations accompany the project page, \url{https://daydreamlive.github.io/DEMON/#experiments}.\par}

\subsection{Per-Frame Control: SDE Source Blending and x0-Target Morph}\label{sec:perframe}

The SDE source-blending control (Section~\ref{sec:engine}) anchors each SDE re-noise step toward the source latents per-frame, with per-frame control over anchoring strength. A per-frame control succeeds if it produces within-generation variation a global scalar cannot, without degrading prompt adherence beyond its intended source-preservation tradeoff; the global do-no-harm quality bar is defined in Section~\ref{sec:quality}. We test three claims: (a) this mechanism is independent of ODE schedule truncation, (b) shaped curves produce per-frame gradients in source preservation that the global scalar denoise does not produce, and (c) text adherence is preserved. Evaluation uses 6 sources spanning electronic, ambient, jazz, lo-fi, metal, and progressive rock, each with a cross-genre cover prompt.

\subsubsection*{Independence from schedule truncation}

Flat SDE curves at matched values produce materially different output from ODE denoise: the 0.5--6 dB waveform SNR divergence between matched SDE and ODE outputs (Table~\ref{tab:sdeclap}) confirms these are fundamentally different operations, not alternative parameterizations of the same effect. SDE flat curves achieve comparable CLAP alignment to corresponding ODE denoise levels (ODE 0.3 = 0.497, SDE flat 0.3 = 0.449; ODE 0.7 = 0.411, SDE flat 0.7 = 0.475) while providing higher spectral similarity to the source (see segmented analysis below), suggesting that step-by-step source anchoring is more effective at source preservation than schedule truncation.

{\footnotesize CLAP computed on a fixed-seed 10s crop (CLAP rand\_trunc) of the 60s output. SNR measures waveform divergence between SDE output and ODE output at the matched denoise value (not quality degradation). Source: techno (electronic), seed 1528.\par}

\begin{table}[!t]
\caption{SDE flat-curve CLAP alignment and divergence from matched ODE denoise.}
\label{tab:sdeclap}
\centering
\footnotesize
\begin{tabular}{@{}lcc@{}}
\toprule
Curve value & CLAP alignment & SNR vs ODE match \\
\midrule
flat 0.3 & 0.449 & 6.1 dB \\
flat 0.5 & 0.480 & 4.7 dB \\
flat 0.7 & 0.475 & 3.4 dB \\
flat 1.0 & 0.425 & 0.5 dB \\
\bottomrule
\end{tabular}
\end{table}

\subsubsection*{Per-frame preservation gradient}

To test whether shaped curves produce genuine per-frame variation (not just a blurred average), we segment each 60-second output into four 15-second windows and measure mel-spectrogram cosine similarity to the corresponding source segment (Table~\ref{tab:gradient}). ODE denoise acts uniformly: the maximum per-segment variation across all ODE denoise levels is 0.027. SDE ramp 0$\rightarrow$1 produces a monotonic descending gradient (0.977 to 0.934, gradient $-$0.043), while ramp 1$\rightarrow$0 produces a steep ascending gradient (0.803 to 0.970, gradient +0.167). To confirm this is reproducible and source-independent, we run the same analysis across 11 seeds and all 6 sources, yielding 66 ramp samples and 66 baseline samples.

\begin{table}[!t]
\caption{Per-segment mel-spectrogram similarity to source (per-frame gradient).}
\label{tab:gradient}
\centering
\scriptsize
\setlength{\tabcolsep}{3pt}
\begin{tabular}{@{}lccccc@{}}
\toprule
Method & \begin{tabular}[b]{@{}c@{}}Seg 1\\(0--15s)\end{tabular} & \begin{tabular}[b]{@{}c@{}}Seg 2\\(15--30s)\end{tabular} & \begin{tabular}[b]{@{}c@{}}Seg 3\\(30--45s)\end{tabular} & \begin{tabular}[b]{@{}c@{}}Seg 4\\(45--60s)\end{tabular} & Gradient \\
\midrule
ODE denoise=0.3 & 0.946 & 0.947 & 0.944 & 0.945 & $-$0.001 \\
ODE denoise=0.7 & 0.900 & 0.916 & 0.914 & 0.919 & +0.018 \\
ODE denoise=1.0 & 0.848 & 0.880 & 0.890 & 0.875 & +0.027 \\
SDE flat 0.5 & 0.969 & 0.969 & 0.962 & 0.965 & $-$0.005 \\
\textbf{SDE ramp 0$\rightarrow$1} & \textbf{0.977} & \textbf{0.971} & \textbf{0.957} & \textbf{0.934} & \textbf{$-$0.043} \\
\textbf{SDE ramp 1$\rightarrow$0} & \textbf{0.803} & \textbf{0.964} & \textbf{0.965} & \textbf{0.970} & \textbf{+0.167} \\
\bottomrule
\end{tabular}
\tnote{Per-segment mel-spectrogram cosine similarity to source (techno, seed 1528). Mel similarity is used rather than waveform SNR because the style-transfer task changes timbre, making waveform comparison uninformative. ODE lines are flat; SDE ramp curves show monotonic gradients.}
\end{table}

The gradient is consistent across all seeds and sources (Table~\ref{tab:gradient-seeds}):

\begin{table}[!t]
\caption{SDE ramp gradient across seeds and sources.}
\label{tab:gradient-seeds}
\centering
\footnotesize
\begin{tabular}{@{}lcc@{}}
\toprule
Configuration & n & Gradient (mean $\pm$ std) \\
\midrule
SDE ramp 0$\rightarrow$1, techno (electronic) & 11 & $-$0.043 $\pm$ 0.002 \\
SDE ramp 0$\rightarrow$1, ambient & 11 & $-$0.134 $\pm$ 0.020 \\
SDE ramp 0$\rightarrow$1, jazz & 11 & $-$0.239 $\pm$ 0.034 \\
SDE ramp 0$\rightarrow$1, lo-fi & 11 & $-$0.399 $\pm$ 0.045 \\
SDE ramp 0$\rightarrow$1, metal & 11 & $-$0.183 $\pm$ 0.057 \\
SDE ramp 0$\rightarrow$1, prog rock & 11 & $-$0.216 $\pm$ 0.019 \\
\bottomrule
\end{tabular}
\tnote{Gradient = segment 4 similarity minus segment 1 similarity. Negative = source preservation decreases over time (expected for ramp 0$\rightarrow$1). 11 seeds per source, 6 sources spanning electronic, ambient, jazz, lo-fi, metal, and progressive rock.}
\end{table}

The ramp gradient is negative for every source and tightly clustered within each source (std 0.002--0.057). The gradient magnitude varies by source (techno: $-$0.043, lo-fi: $-$0.399), reflecting how much the cross-genre cover task transforms each source, but the direction is universally consistent. When compared to the corresponding SDE baseline (flat 1.0) gradients, the ramp always pushes the gradient more negative: for example, the metal baseline (flat 1.0) has a strong positive gradient (+0.324) that the ramp overcomes entirely, producing a large net swing. This confirms the SDE curve produces genuine per-frame control independent of source material.

The ramp 1$\rightarrow$0 saturates after segment 2 rather than continuing to rise linearly. This reflects the DiT's global temporal attention: frames with high curve values (less preservation) influence the velocity predictions for neighboring frames, smoothing the per-frame control. The control operates per-frame, but the model's response is coupled through attention.

The companion project page presents the curve direction studied here as a single curated variant, the asymptotic $1 - t^3$ shape (same high$\rightarrow$low orientation as ramp 1$\rightarrow$0, steeper at the end), with audible examples on two source fixtures. See \url{https://daydreamlive.github.io/DEMON/#experiments}.

\subsubsection*{Text adherence}

Shaped SDE curves show whole-file CLAP alignment (ramp 0$\rightarrow$1 = 0.519, ramp 1$\rightarrow$0 = 0.489) within the range of flat curves (0.425--0.480), consistent with source-preserving frames pulling the global score toward the source rather than the prompt. This is the intended tradeoff: Section~\ref{sec:quality} confirms that the SDE source-blending curve produces a negative CLAP delta proportional to its source anchoring.

\subsubsection*{x0-Target Morph: Per-Frame Control Toward an Independent Target}

The x0-target morph (Section~\ref{sec:pipeline}) blends toward an independent precomputed target latent B rather than the source, so we measure per-segment similarity to the \emph{target} B, not the source. Segmenting each morph into four equal windows ($\sim$14\,s each) and measuring mel-spectrogram cosine similarity to the corresponding window of B, a flat $\alpha{=}0$ baseline (pure cover-of-A) shows no gradient toward B, a flat $\alpha{=}1$ collapses every segment onto B, and a shaped $\alpha$ ramp $0\rightarrow1$ drives a monotone climb toward B (Table~\ref{tab:x0morph}). The direction holds across 3 source$\rightarrow$target pairs $\times$ 3 seeds ($n{=}9$): mean gradient $+0.217 \pm 0.051$ (min $+0.137$), every sample monotone toward B. Swap-tick and steady-state outputs pass the same objective probes as migration (no NaN/Inf, peak amplitude within 1\% of reference, high-frequency energy within the endpoints' band); consistent with Section~\ref{sec:discussion} we report this as objective-probe-clean, not a proven perceptual result. The live onset of a \texttt{set\_shared\_curve} write on \texttt{x0\_target\_strength} is reported with the other shared-mutable controls in Table~\ref{tab:propagation}.

\begin{table}[!t]
\caption{x0-target morph: per-segment mel-spectrogram similarity to the target latent B.}
\label{tab:x0morph}
\centering
\footnotesize
\begin{tabular}{@{}lccccc@{}}
\toprule
$\alpha$ schedule & Seg 1 & Seg 2 & Seg 3 & Seg 4 & Gradient \\
\midrule
flat $0$ (cover-of-A) & 0.765 & 0.736 & 0.829 & 0.709 & $-0.056$ \\
\textbf{ramp $0\rightarrow1$} & \textbf{0.789} & \textbf{0.908} & \textbf{0.998} & \textbf{1.000} & \textbf{$+0.211$} \\
flat $1$ (full B) & 1.000 & 1.000 & 1.000 & 1.000 & $+0.000$ \\
\bottomrule
\end{tabular}
\tnote{Per-segment mel-spectrogram cosine similarity to the morph target B (four equal $\sim$14.2\,s windows). Gradient = segment 4 minus segment 1. Highlighted pair: deathcore$\rightarrow$ambient cover, seed 1528; direction confirmed across the 3$\times$3 matrix in the text. Similarity is to the \emph{target} B, distinguishing this from the source-similarity measurement of Table~\ref{tab:gradient}.}
\end{table}

\subsection{Quality Evaluation}\label{sec:quality}

Our contributions are infrastructure (streaming pipeline, windowed VAE, TRT acceleration) and a per-frame source-blending control on the SDE step. We do not claim to improve the base model's generation quality. The relevant quality question is therefore: \emph{do our additions degrade the output compared to unmodified batch-mode generation?}

We use CLAP text-audio alignment \cite{b9} (the \texttt{laion/larger\_clap\_music\_and\_speech} checkpoint) to answer this. CLAP maps audio and text into a shared embedding space; cosine similarity measures how well the audio matches its prompt. Cover tasks inherently balance source preservation against prompt adherence, so the metric of interest is the \emph{delta} between each variant and the no-modification baseline.

\subsubsection*{Stream vs Batch Parity}
We compare stream pipeline output (depth=8, 8 denoising steps) to batch-mode generation (8-step sequential loop) at matched seeds and conditions (Table~\ref{tab:parity}).

\begin{table}[!t]
\caption{Stream-versus-batch parity.}
\label{tab:parity}
\centering
\footnotesize
\begin{tabular}{@{}lccp{0.24\columnwidth}@{}}
\toprule
Method & \begin{tabular}[b]{@{}c@{}}CLAP\\alignment\end{tabular} & \begin{tabular}[b]{@{}c@{}}CLAP\\delta\end{tabular} & SNR vs batch (dB) \\
\midrule
Batch (reference) & 0.246 & ref & ref \\
Stream (depth=8) & 0.246 & 0.000 & $\infty$ (sample-identical, 16-bit PCM) \\
\bottomrule
\end{tabular}
\end{table}

The streaming pipeline produces output sample-for-sample identical to batch-mode generation at the 16-bit PCM render (infinite SNR, i.e.\ zero per-sample difference, with identical CLAP and RMS) across all 6 evaluation sources. The two paths run the same deterministic math, so the rendered audio is identical to its quantization precision; we report this at the audio level (a $\geq{\sim}90$ dB floor) and do not separately assert float- or latent-level bit-equality, which batched-versus-sequential GPU kernel differences could perturb below the PCM floor. Either way, the streaming infrastructure adds no audible quality cost.

\subsubsection*{SDE Curve Ablation}
We measure CLAP delta vs.\ matched baseline (same source, seed, no curve) across 6 sources and 10 seeds at a single operating point (denoise=1.0, cover task) because SDE operates in its own fixed-denoise regime independent of the ODE operating-point variations (Table~\ref{tab:sde-ablation}):

\begin{table}[!t]
\caption{SDE curve CLAP-delta ablation against matched baseline.}
\label{tab:sde-ablation}
\centering
\footnotesize
\begin{tabular}{@{}lccc@{}}
\toprule
Curve & n & \begin{tabular}[b]{@{}c@{}}CLAP delta\\(mean $\pm$ std)\end{tabular} & \begin{tabular}[b]{@{}c@{}}SNR vs\\baseline (dB)\end{tabular} \\
\midrule
sde\_denoise\_curve (ramp 0--1) & 47 & $-$0.073 $\pm$ 0.115 & $-$1.1 \\
\bottomrule
\end{tabular}
\tnote{Expected n=60 (6 sources $\times$ 10 seeds); 13 samples excluded due to source-level evaluation failures.}
\end{table}

The negative CLAP delta is consistent with source preservation reducing prompt adherence, as expected for a curve that continuously anchors toward source latents at every solver step. This confirms the source-preservation mechanism is working: the output is trading prompt adherence for source fidelity at frames where the curve value is low.

\subsubsection*{Distributional Quality (FAD)}
We compute Frechet Audio Distance \cite{b10}, using CLAP embeddings (the same checkpoint) in place of the original VGGish features, against a 500-track subset of FMA-small \cite{b24} as a distributional sanity check. The reference set is genre-mismatched (FMA spans 8 genres; our samples are covers of 6 sources), so absolute scores reflect style distance rather than quality. The relative scores confirm that the SDE curve does not degrade output beyond its intended tradeoff: baseline FAD = 0.649 (n=240), sde\_denoise = 0.829 (n=60). The sde\_denoise elevation is expected, as source-anchoring pulls output toward the input recording rather than the model's learned distribution.

{\footnotesize FAD uses n=60 (full 6 sources $\times$ 10 seeds) because it compares against the FMA reference distribution rather than requiring per-source matched pairs; the matched-baseline CLAP delta above is restricted to the n=47 pairs that survived the source-level matching filter.\par}

\subsection{Parameter-Change Propagation}\label{sec:propagation}

\subsubsection{Parameter-Change Latency}\label{sec:paramlatency}

How quickly a parameter change becomes audible is set by its propagation class (Section~\ref{sec:pipeline}). Table~\ref{tab:propagation} measures the per-request, per-step shared-mutable, and model-weight classes at depth 8 (the denoise row is its per-request, heterogeneous-mode behavior; the migrated-schedule class is characterized in Section~\ref{sec:pipeline}).

We use depth 8 as the reference: D=S=8 gives the cleanest step structure (completion interval equals tick, no burst phase offset; depth 4 is the production operating point, Section~\ref{sec:depth}). All five per-request rows are a clean step at tick S, bit-identical to the pre-change reference until the first post-change slot completes its schedule (a slot freezes its request at \texttt{submit()}, Section~\ref{sec:pipeline}). The denoise switch produces the largest latent-space step (RMS 1.258), confirming denoise is the most impactful single parameter; the prompt, hint, source-audio, and timbre-reference changes span a comparable 0.94--1.21 band, all substantial steps sharing the single tick-8 ($\sim$648\,ms) drain floor rather than splitting into fast and slow.

Shared mutable curves bypass the drain. Setting the SDE curve from 0.1 to 0.95 via shared state takes effect on the very next completion (tick 1) and builds progressively to plateau by tick 5, rather than waiting for tick 8. This is the architectural distinction that matters for interactive use: per-step parameters (Section~\ref{sec:pipeline}) react on the first tick that runs the solver loop, while per-request parameters bound at submit time wait for ring-buffer drain. The progressive shape (rather than a single-tick jump) reflects the burst of in-flight slots each contributing the new curve to a different number of remaining steps. The x0-target morph shows the same signature, converging over a four-step ramp set by its refinement-half gate (Sections~\ref{sec:pipeline}, \ref{sec:perframe}; Table~\ref{tab:propagation}).

\begin{table*}[!t]
\caption{Parameter-change propagation at depth 8.}
\label{tab:propagation}
\centering
\footnotesize
\begin{tabular}{@{}lllp{0.10\textwidth}p{0.30\textwidth}@{}}
\toprule
Parameter change & First effect & Converged & First-effect RMS & Propagation \\
\midrule
Denoise (1.0$\rightarrow$0.5) & tick 8 / 648ms & tick 8 / 648ms & 1.258 & Step function at tick S \\
Prompt switch & tick 8 / 648ms & tick 8 / 648ms & 0.941 & Step function at tick S \\
Hint strength (1$\rightarrow$0) & tick 8 / 648ms & tick 8 / 648ms & 1.210 & Step function at tick S \\
Source audio swap & tick 8 / 648ms & tick 8 / 648ms & 1.095 & Step function at tick S \\
Timbre reference (1$\rightarrow$0) & tick 8 / 648ms & tick 8 / 648ms & 1.163 & Step function at tick S \\
SDE curve (shared) & tick 1 / 81ms & tick 5 / 405ms (plateau) & 0.64 (tick 1) & Progressive (tick 1: 0.64; tick 5: 1.25) \\
x0\_target\_strength (shared) & tick 1 / 81ms & tick 4 / 324ms & 0.79 (tick 1) & Progressive (tick 1: 0.79; tick 4: 0.97 plateau) \\
LoRA refit & \multicolumn{2}{l}{1 tick / $\sim$81ms (any depth) + 1.2s refit} & & Immediate (model weights) \\
\bottomrule
\end{tabular}
\tnote{All tests at depth=8, RTX 5090. Source swap uses denoise=0.7 (source latents are only mixed at denoise$<$1.0); other tests at denoise=1.0 except the denoise-switch row itself, which goes 1.0$\rightarrow$0.5. RMS measured in latent space against the last pre-change completion. All five per-request rows hit the same tick-8 drain floor; the 648\,ms figure is 8$\times$ the 81.1\,ms depth-8 tick (Table~\ref{tab:depth}), and the hint/source/timbre re-measurement run measured 82--84\,ms ticks (within clock-state variance, so the faster canonical figure is reported). Timbre reference has no scalar knob: it is a \texttt{refer\_latent} blend between silence and the source, re-encoded, so it propagates as a conditioning swap (the prompt class). LoRA refit time is a one-time cost; post-refit inference is immediate. SDE shared-curve switch values (0$\rightarrow$0.64$\rightarrow$0.96$\rightarrow$1.13$\rightarrow$1.20$\rightarrow$1.25 plateau over ticks 0--5) verified GPU-independent across 5090, 4090, and 3090 in Section~\ref{sec:crossgpu}. The x0\_target\_strength row is from the morph evidence run; ticks count denoising passes after the live set (the convention of Section~\ref{sec:depth}). The ring returns each completion one pass after its final step, so the raw returned-RMS series leads with the unchanged in-flight completion (0) before 0.79, 0.91, 0.96, 0.97; that leading zero is the pipeline's fixed one-completion output latency, identical across control classes. The four-pass ramp reflects the refinement-half gate (steps 4--7).}
\end{table*}

\subsubsection{Heterogeneous Scheduling vs.\ Global Reset Ablation}\label{sec:ablation}

Section~\ref{sec:paramlatency} measured how fast a denoise change reaches the output (S ticks at depth=8). That measurement is silent on what happens to the rest of the queue \emph{during} those S ticks. Section~\ref{sec:pipeline} claimed that per-slot heterogeneous scheduling lets in-flight slots continue to their natural completion on the schedules they were born with, so the output stream is uninterrupted across the change. The natural counterfactual is the upstream StreamDiffusion design, whose global-timestep \texttt{prepare()} rebuild wipes the in-flight queue on every strength change (Section~\ref{sec:pipeline}). We test the two regimes directly.

The global-reset baseline is emulated by clearing the slot array (\texttt{pipeline.\_slots = [None] * depth}) whenever the submitted denoise value differs from the previous tick. This reproduces the observable behavior of \texttt{prepare()} (every in-flight generation is discarded and the buffer must re-warmup for D ticks) without reimplementing the upstream queue-rebuild path. All other engine state (TRT context, schedule cache, compiled graphs) is untouched, so the ablation isolates the slot-state-versus-global-state design choice. We run three arms at depth=8, S=8, on the 5090 (Table~\ref{tab:het-ablation}).

\textbf{Mechanism.} A 1.0$\rightarrow$0.5 denoise switch is applied after warmup, and the per-slot \texttt{(denoise, step\_idx)} tuple is snapshotted after every tick. In per-slot mode the slot array carries both denoise values simultaneously for ticks 0--6 (the drain window: old slots finishing their born-with schedules alongside newly admitted denoise=0.5 slots), and collapses to \texttt{\{0.5\}} at tick 7 once the last old slot retires. In global-reset mode the slot array is \texttt{\{0.5\}} from tick 0 onward by construction. The batched timestep tensor consumed by the DiT is therefore heterogeneous in the per-slot regime (different rows are at different denoise strengths in the same forward pass) and uniform in the global-reset regime.

\textbf{Single-switch throughput.} Across the 24-tick post-change window, per-slot scheduling produces a completion on every tick (24/24) with rms=0 against the last pre-change reference for ticks 0--7 (old slots draining at the old denoise, output identical to pre-change) and a step to rms=1.234 at tick 8 (first new-denoise completion). The global-reset baseline produces zero completions for ticks 0--7 (the depth-tick refill window) and 16 completions for ticks 8--23, with first new-denoise completion also at tick 8. Both modes hit the same S-tick architectural floor for parameter-change latency; the modes differ only in what happens during the drain.

\textbf{Continuous slider sweep.} A 60-tick linear sweep of denoise from 1.0$\rightarrow$0.5$\rightarrow$1.0 with one slider step per tick is the stress test for the per-slot mechanism: under global-reset every tick wipes the queue. Per-slot scheduling sustains a 100\% completion rate (60/60) at a mean tick latency of 81.1 ms, indistinguishable from the steady-state held-denoise tick. The global-reset baseline collapses to a 1.7\% completion rate (1/60, only the warmup-state tick before the first slider movement); ticks have a mean latency of 15.8 ms because the buffer is almost always empty, but no useful output is produced. The cheap ``fast'' tick under global-reset is the absence of work.

\begin{table*}[!t]
\caption{Heterogeneous scheduling versus global-reset ablation.}
\label{tab:het-ablation}
\centering
\footnotesize
\begin{tabular}{@{}p{0.44\textwidth}p{0.15\textwidth}p{0.27\textwidth}@{}}
\toprule
Metric & Per-slot (DEMON) & Global-reset (StreamDiffusion-style) \\
\midrule
Heterogeneity during drain (ticks 0--6 post-switch) & \{0.5, 1.0\} coexist & \{0.5\} only (queue wiped) \\
Completions during 24-tick post-switch window & 24/24 & 16/24 \\
Consecutive dead-air ticks after change & 0 & 8 ($\sim$649 ms) \\
First-new-effect tick & tick 8 & tick 8 \\
Continuous slider sweep (60 ticks, denoise every tick) completion rate & 60/60 (100\%) & 1/60 (1.7\%) \\
Sweep mean tick latency & 81.1 ms & 15.8 ms (empty-buffer ticks) \\
\bottomrule
\end{tabular}
\tnote{Depth=8, S=8, seed=1528, RTX 5090. Single-switch arm reproduces the 5.8.1 1.0$\rightarrow$0.5 denoise row; the contribution here is the per-tick completion and slot-state trace across the drain. The global-reset emulation clears the slot array on denoise change; all other engine state is preserved, so the per-tick latency in global-reset mode reflects the cost of running the forward on a partially-filled buffer, not extra rebuild work.}
\end{table*}

Per-slot heterogeneous scheduling does not reduce parameter-change latency: that floor is set by the denoising schedule itself and matches the global-reset baseline at tick 8 either way. The contribution is throughput preservation across the drain. Under a single discrete change this manifests as an 8-tick dead-air gap ($\sim$649 ms at depth=8) avoided; under continuous slider movement it is the difference between a usable live control surface and one that produces audio 1.7\% of the time. The mechanism is also a property of the streaming-diffusion code path, not the hardware: the cross-GPU runs in Section~\ref{sec:crossgpu} confirm the architectural patterns of Section~\ref{sec:paramlatency} hold within 0.02 RMS across 5090, 4090, and 3090; we did not re-run this ablation across GPUs because the wipe-versus-no-wipe behavior cannot vary with hardware.

\textbf{Which baseline this is, and which it is not.} The global-reset arm is the \emph{queue-wiping} design: the original StreamDiffusion \texttt{prepare()} path, which rebuilds the global timestep tensor and zeros the buffer on every strength change. It is not the strongest prior baseline. An \emph{in-place-edit} design, such as the daydreamlive fork \cite{b23}, rewrites the global timestep list under the buffer without zeroing it, so it keeps the buffer full and would sustain a completion rate near 100\% under the same sweep. That in-place global edit is exactly the in-flight schedule migration of Section~\ref{sec:pipeline} expressed on a single shared schedule, which holds completion under a continuous sweep. So the 100\%-versus-1.7\% gap measured here is specifically against the wipe design; against the stronger in-place-edit baseline, completion rate is a tie at $\sim$100\% and the per-slot advantage is not throughput but \emph{trajectory coherence}: every per-slot completion is a valid single-schedule trajectory, whereas the in-place global edit produces the trajectory-incoherent hybrids characterized in Section~\ref{sec:pipeline}. To our knowledge this is also the first reported measurement of per-slot heterogeneous scheduling against either baseline; the fork was never written up or benchmarked.

An illustrative audio analog of the contrast above (2 s of denoise=1.0 output joined to 2 s of denoise=0.5 output via either a 50 ms equal-power crossfade or $\sim$648 ms of silence) is available on the companion project page, \url{https://daydreamlive.github.io/DEMON/#experiments}.

\subsubsection{Pipeline Depth Tradeoff}\label{sec:depth}

Pipeline depth trades generation throughput against parameter-change responsiveness; all depths share the same 8 denoising steps at identical quality (Section~\ref{sec:pipeline}), so depth sets only the in-flight count. One completion emerges every S/D ticks, and because GPU batching is sub-linear (B=8 costs $\sim$6.1$\times$ B=1, Section~\ref{sec:complatency}), higher depth yields more generations per second despite a more expensive tick. When D $<$ S, completions arrive in bursts of D followed by (S$-$D) dry ticks, so per-request convergence depends on burst-cycle position. (The submit queue is capped at D to bound backlog at the 1/tick submit rate.)

\begin{table*}[!t]
\caption{Pipeline depth tradeoff.}
\label{tab:depth}
\centering
\footnotesize
\begin{tabular}{@{}cccccc@{}}
\toprule
Depth & Tick (ms) & Completion interval (ms) & Gens/sec & Denoise first-effect tick & Denoise first-effect (ms) \\
\midrule
1 & 14.0 & 112.0 & 8.9 & tick 8 & 112 ms \\
2 & 24.3 & 97.2 & 10.3 & tick 9 & 219 ms \\
4 & 42.8 & 88.5 & 11.3 & tick 11 & 471 ms \\
8 & 81.1 & 81.1 & 12.3 & tick 8 & 649 ms \\
\bottomrule
\end{tabular}
\tnote{8 denoising steps at all depths. denoise=1.0, 60s generation, RTX 5090. First-effect tick is for a denoise 1.0$\rightarrow$0.5 switch, measured directly via per-tick latent-RMS vs the last pre-change completion (RMS = 0.0 for ticks 0\ldots k$-$1, jumping to $\sim$1.24 at tick k); a prompt switch gives identical first-effect ticks (8/9/11/8 at depths 1/2/4/8), confirming per-request-class behavior. First-effect ms uses the Tick column; the denoise sweep's own per-run ticks (15.7/25.9/43.7\,ms at depths 1/2/4) match it within clock-state variance. Exact first-effect tick depends on the slot phase at the moment of change (the burst-cycle position can shift it by $\pm$1--3 ticks); single-position measurement reported. LoRA refit, by contrast, takes effect on the very next tick at any depth (14.0--81.1 ms) because it modifies shared decoder weights rather than admitting a new slot. The Tick column is the bare batched-forward latency (and the per-step onset); Completion interval and Gens/sec are observed over a separate 200-tick throughput run. Depths 1, 2, and 8 land on S$\cdot$tick/D exactly (112.0, 97.2, 81.1 ms); the depth-4 interval (88.5 ms) sits just above 2$\times$tick (85.6 ms), the residual being run-to-run variance between the two captures.}
\end{table*}

Throughput rises 38\% from depth 1 to 8 (8.9$\rightarrow$12.3 gens/sec) while per-request first-effect rises 5.8$\times$ (112$\rightarrow$649 ms), since each new-conditioning slot must still run its full S steps. Depth 1 (112 ms first-effect, 8.9 gens/sec) and depth 8 (649 ms, 12.3 gens/sec) bound the tradeoff; we run depth 4 as the sweet spot: 11.3 gens/sec (within 8\% of peak), an 88.5 ms completion interval (within 9\% of depth 8's smoothness), and a measured 471 ms first-effect, 178 ms below depth 8 (its theoretical floor is 342 ms = 8$\times$42.8 ms; the burst-cycle phase offset adds the rest, see Table~\ref{tab:depth} caption). Per-step parameters are exempt: the shared-mutable curve is read every step, so it onsets on the next tick at any depth (42.8 ms at depth 4, 81 ms at depth 8, Section~\ref{sec:pipeline}), keeping the dominant interaction surface responsive regardless of depth.

\subsection{Cross-GPU Generalization}\label{sec:crossgpu}

To verify that the streaming architecture's behavior is intrinsic and not specific to the 5090's hardware characteristics, we ran the full bench suite (component latency, depth sweep, end-to-end streaming, parameter-change propagation, windowed VAE, full-pipeline cover) on RTX 4090 and RTX 3090 instances provisioned on Vast.ai. All three GPUs use the same TRT engine build pipeline (mixed precision, batch\_max=8, seq\_max=1500) and the same bench harness; only the host hardware and driver differ (Table~\ref{tab:crossgpu}).

\begin{table}[!t]
\caption{Cross-GPU operating points.}
\label{tab:crossgpu}
\centering
\footnotesize
\begin{tabular}{@{}p{0.42\columnwidth}ccc@{}}
\toprule
Operating point & 5090 & 4090 & 3090 \\
\midrule
Decoder B=8 (60s, ms) & 80.6 & 108.4 & 234.4 \\
VAE decode full 60s (ms) & 55.9 & 80.3 & 141.5 \\
VAE decode 15s window (ms) & 20.0 & 25.9 & 43.7 \\
Streaming tick depth=8 (ms) & 81.1 & 112.6 & 240.2 \\
End-to-end depth=8 (tick + 15s windowed decode, ms) & 102 & 138.9 & 286.7 \\
Throughput depth=8 (gens/sec) & 12.3 & 8.9 & 4.2 \\
Working-set VRAM (MB) & n/a & 12273 & 12228 \\
\bottomrule
\end{tabular}
\tnote{5090 numbers from Tables~\ref{tab:complatency} and~\ref{tab:propagation}. 4090 driver 580.126.09, 3090 driver 580.82.09; both ran the default cu13 TRT path without patching. 5090 working-set VRAM was not captured in this run. 5090 prompt-switch and SDE-shared-curve measurements come from a separate run on the same engine binaries (\texttt{output/prompt\_switch\_5090.json}, \texttt{output/sde\_shared\_5090.json}); 4090 and 3090 measurements come from the cross-GPU bench harness.}
\end{table}

The 4090 reaches 8.9 generations per second at depth 8 (above real-time for 60-second outputs); the 3090 reaches 4.2 generations per second (sub-real-time but functional). Working-set VRAM is approximately 12 GB on both consumer cards, comfortably within their 24 GB capacity, indicating the architecture should run on 16 GB cards as well.

\textbf{Windowed VAE speedup is GPU-independent.} The 15s window vs.\ full 60s decode speedup is 2.80x on the 5090, 3.10x on the 4090, and 3.22x on the 3090, confirming that the receptive-field-based optimization (Section~\ref{sec:vae}) reflects a property of the Oobleck VAE rather than a GPU-specific effect.

\textbf{Parameter-change propagation is GPU-independent.} Repeating the prompt-switch test (Section~\ref{sec:propagation}) on the 4090 and 3090 produces the same propagation pattern at the same per-tick latent magnitudes: post-switch RMS in the latent space is zero for ticks 0--7 (old-conditioning slots draining) and $\sim$0.93 starting at tick 8 (new-conditioning slots arriving), within 0.02 of the 5090's 0.94. This confirms that the S-tick step-function pattern in Table~\ref{tab:propagation} is an architectural property of the ring buffer rather than a 5090-specific timing artifact.

The same per-tick capture for the shared SDE curve (Section~\ref{sec:pipeline}) gives the contrasting pattern: instead of zero-then-step at tick 8, every GPU shows immediate effect at tick 1 with progressive convergence over the next several ticks. The 5090, 4090, and 3090 again sit within 0.01 of each other at every tick.

The shared-state write itself is sub-millisecond on all three GPUs (5090 0.005 ms, 4090 0.062 ms, 3090 0.059 ms); cost is dominated by the first tick that consumes the new value. Both signatures (per-request zero-then-step at tick 8, shared-mutable tick-1 onset) are thus a property of the ring-buffer architecture, not 5090-specific timing; per-tick tables for both are in the supplementary release.

\section{Discussion}\label{sec:discussion}

The whole system answers to one aim, an instrument a musician can play, which needs a control surface that is both broad (it exposes many parameters and shapes them per-frame) and responsive (controls heard quickly enough to perform with). On the responsiveness pillar, the central result is that sufficient diffusion throughput turns denoising parameters into live performance controls, but the ring-buffer pipeline that enables this throughput propagates per-request changes only at its S-tick drain floor. Our central response for the most impactful control, scalar denoise, is per-slot heterogeneous scheduling: each slot owns its timestep schedule, so a moving slider is tracked without wiping in-flight state and every output is a coherent single-schedule trajectory (Section~\ref{sec:pipeline}), where the upstream global-reset design produces audio on 1.7\% of ticks under the same sweep (Section~\ref{sec:ablation}). A separate infrastructure mechanism handles a different parameter class: any value consulted at every solver step (the per-frame curves of Section~\ref{sec:engine}) can be held as shared mutable state and take effect on the next tick (Section~\ref{sec:pipeline}); the per-frame source-blending control on the SDE step is one instance, a framewise axis that complements scalar denoise (Section~\ref{sec:perframe} confirms it acts independently of schedule truncation), while the x0-target morph is the worked example (Section~\ref{sec:pipeline}). Applying the same shared-mutable pattern to the structural schedule gives in-flight schedule migration, a 1-tick-onset option for denoise (Section~\ref{sec:pipeline}). Together these populate four parameter-propagation classes separated by onset and convergence latency. The pipeline depth tradeoff (Section~\ref{sec:depth}) governs per-request responsiveness: depth=1 gives 112 ms convergence at 8.9 gens/sec and depth=8 gives 649 ms at 12.3 gens/sec, with depth 4 the recommended balance we run in practice (471 ms measured at 11.3 gens/sec, within 8\% of the peak throughput). At every depth the shared-mutable curves (and, with the caveat above, migration) give 1-tick onset (42.8 ms at depth 4, 81 ms at depth 8), so per-step control stays responsive regardless of the chosen depth.

\subsection*{Limitations}

The system inherits ACE-Step 1.5's limitations in music quality and genre coverage. More fundamental are the limits intrinsic to streaming diffusion as a control surface. The control is over denoising \emph{dynamics}, not musical \emph{content}: a per-frame curve shapes how a region transforms but cannot place a note, enforce a chord, or guarantee a downbeat, which follow from the base model's learned distribution. The throughput that makes controls live is also in direct tension with per-request onset: higher pipeline depth raises generations per second but pushes per-request changes further behind the S-tick drain floor, and the shared-mutable escape bypasses that floor only for parameters read at every solver step. Structural parameters (step count, conditioning and source identity) cannot be made per-step without breaking trajectory coherence, so they remain permanently bound to that floor. Even the fast per-step controls carry a one-tick onset floor of a single forward pass, which grows with sequence length (the depth-8 tick rises from 81~ms at 60~s to $\sim$381~ms at 240~s, Section~\ref{sec:e2e}). The curves also apply independently at 25~Hz, but the DiT's global temporal attention couples the model's response across frames, so sharp control transitions produce smoothed output transitions (Section~\ref{sec:perframe}); truly independent per-frame control is therefore unattainable in any globally-attentive model. LoRA refit stores merged weights at runtime, raising VRAM by $\sim$40\% when a LoRA is active, and DiT working-set VRAM scales with song length, which a larger card mitigates and further quantization could reduce.

A second class of limits is one of scope rather than mechanism. The system generates fixed-length songs, not an endless evolving stream: ACE-Step 1.5 correctly opens and closes a piece as generation approaches the bounds of its latent, so the natural fit is remixing a whole song with an intended beginning and end, or working on loops, rather than the indefinite continuation that autoregressive models such as Lyria \cite{b20} provide by construction. Reaching open-ended generation would require fine-tuning or a different formulation. Source conditioning is also static and single-source: the input latent is encoded once up front, so the system neither accepts live real-time audio input nor re-encodes a new latent per frame, both of which await integrating a streaming on-device autoencoder. Finally, the reference-song structure guidance inherited from ACE-Step, while useful, is broad and imperfect; precise beat- or melody-level preservation would need finer conditioning, such as a ControlNet-style \cite{b15} adapter. All three remain open directions.

A further limitation is evaluative. All quality and parity evidence in this paper is objective: CLAP text-audio alignment, FAD, and waveform/latent SNR. We conduct no listening test, MOS, or other human perceptual study, so the claims that motivate the system, that the throughput makes denoising parameters usable as live performance controls, that the response is fast enough to function as a musical control surface, and that the resulting instrument is musically compelling to play, rest on latency measurements and objective proxies rather than subjective listener data. This applies in particular to in-flight schedule migration (Section~\ref{sec:pipeline}): its swap-tick hybrids pass our objective probes (no NaN, amplitude and high-frequency energy within the steady-state band), but our own listening finds an audible difference, the faster onset carries a transient muffling until the buffer drains, so the transition is perceptibly less smooth than the coherent-trajectory heterogeneous mode and we have no formal perceptual data to quantify it. This is why heterogeneous scheduling, not migration, is our denoise control surface, with migration offered as an explicit fast-onset option. A perceptual evaluation with musicians is the most important piece of validation left to future work.

\subsection*{Active Research Program}

The engine described in this paper is the foundation of a broader research program in real-time controllable music diffusion. Two extensions are under active development. Companion technical notes for both extensions are available on the project page, \url{https://daydreamlive.github.io/DEMON/}.

\textbf{Distilled VAE decoder.} A FastOobleckDecoder (51.7M parameters, 61\% of teacher) trained with L1 + multi-resolution STFT + feature distillation losses, targeting further reduction of the VAE decode bottleneck beyond windowed decode.

\textbf{Latent channel semantics.} Empirical characterization of ACE-Step 1.5's 64-channel VAE latent space, mapping channels to perceptual effects for targeted channel-level manipulation during diffusion.

\textbf{ACE-Step XL.} The same build path extends to the larger ACE-Step XL model (W8A8-quantized DiT backbone, same fp32-AdaLN/RMSNorm rule), transparent to the ring-buffer machinery; no XL benchmarks are reported here, and characterization is left to future work.

\section{Conclusion}\label{sec:conclusion}

DEMON makes diffusion denoising playable as a real-time musical instrument: a control surface that is broad, the per-frame curves of Section~\ref{sec:engine} shaping many parameters at 25~Hz, and as responsive as each control's place in the denoising loop allows. Sufficient throughput turns the denoising parameters into live controls; the ring-buffer pipeline that enables it then holds per-request changes at an S-tick drain floor. Our mechanisms (per-slot heterogeneous scheduling for scalar denoise, shared-mutable per-step state for parameters read at every solver step, and in-flight schedule migration) separate streaming-diffusion parameters into four propagation classes by onset and convergence latency: per-request (S-tick onset and convergence), migrated schedule (1-tick onset, S-tick convergence, via transient hybrids), per-step shared-mutable (1-tick onset, sub-S convergence), and model weights (immediate). The enabling infrastructure (an audio streaming pipeline, windowed VAE decode, and TensorRT mixed-precision engines) reaches an 81 ms decoder tick at depth 8 (12.3 decoder completions per second; $\sim$88 ms and 11.4/s end-to-end with the windowed VAE), with depth 4 the operating point we run in practice (11.3 generations per second at a measured $\sim$470 ms per-request first-effect, 178 ms below depth 8's 649 ms).

\section*{Acknowledgments}
The author thanks Qiang Han and Tom Neumann for technical feedback and review; Rafal Leszko, Marco Tundo, Gioele Cerati, and Hunter Hillman for their work on the demo; and the Daydream team and friends who tested early builds.


\end{document}